# Very Large Telescope Observations of Interstellar Comet 3I/ATLAS. II. From Quiescence to Glow: Dramatic Rise of Ni I Emission and Incipient CN Outgassing at Large Heliocentric Distances*

Rohan Rahatgaonkar[1], Juan Pablo Carvajal[1], Thomas H. Puzia[1], Baltasar Luco[1], Emmanuel Jehin[2], Damien Hutsemékers[2], Cyrielle Opitom[3], Jean Manfroid[2], K. Aravind[2], Michaël Marsset[4], Bin Yang[5], Laura Buchanan[6,7], Wesley C. Fraser[6,7], John Forbes[8], Michele Bannister[8], Dennis Bodewits[9], Bryce T. Bolin[10], Matthew Belyakov[11], Matthew M. Knight[12], Colin Snodgrass[3], Erica Bufanda[3], Rosemary Dorsey[13], Léa Ferellec[3], Fiorangela La Forgia[14], Manuela Lippi[15], Brian Murphy[3], Prasanta K. Nayak[1,16], and Mathieu Vander Donckt[2]

[1] Institute of Astrophysics, Pontificia Universidad Católica de Chile, Av. Vicuña Mackenna 4860, 7820436 Macul, Santiago, Chile; rrohan@uc.cl
[2] STAR Institute, University of Liège, Allée du 6 août, 19, 4000 Liège (Sart-Tilman), Belgium
[3] Institute for Astronomy, University of Edinburgh, Royal Observatory, Edinburgh, EH9 3HJ, UK
[4] European Southern Observatory (ESO), Alonso de Cordova 3107, 1900, Casilla Vitacura, Santiago, Chile
[5] Instituto de Estudios Astrofísicos, Facultad de Ingeniería y Ciencias, Universidad Diego Portales, 8370191, Santiago, Chile
[6] National Research Council of Canada, Herzberg Astronomy and Astrophysics Research Centre, 5071 W. Saanich Road, Victoria, BC V9E 2E7, Canada
[7] Department of Physics and Astronomy, University of Victoria, Elliott Building, 3800 Finnerty Road, Victoria, BC V8P 5C2, Canada
[8] School of Physical and Chemical Sciences—Te Kura Matū, University of Canterbury, Private Bag 4800, Christchurch 8140, New Zealand
[9] Physics Department, Edmund C. Leach Science Center, Auburn University, Auburn, AL 36849, USA
[10] Eureka Scientific, Oakland, CA 94602, USA
[11] Division of Geological and Planetary Sciences, California Institute of Technology, Pasadena, CA 91125, USA
[12] Physics Department, United States Naval Academy, 572C Holloway Road, Annapolis, MD 21402, USA
[13] Department of Physics, University of Helsinki, P.O. Box 64, 00014 Helsinki, Finland
[14] Dipartimento di Fisica e Astronomia-Padova University, Vicolo dell'Osservatorio 3, 35122 Padova, Italy
[15] INAF, Osservatorio Astrofisico di Arcetri, Largo E. Fermi 5, 50125 Firenze, Italy

## Abstract

We report Very Large Telescope spectroscopy of the interstellar comet 3I/Asteroid Terrestrial-impact Last Alert System (C/2025 N1), from $r_h \simeq 4.4$ to 2.85 au, using X-Shooter (300–550 nm, $R \simeq 3000$) and the Ultraviolet and Visual Echelle Spectrograph (optical, $R \simeq 35$–80 k). The coma is dust-dominated, with a fairly constant red optical continuum slope (∼21%–22%/1000 Å). We report the detection of CN emission and also detect numerous Ni I lines, while Fe I remains undetected, potentially implying efficiently released gas-phase Ni. At $r_h \simeq 3.14$ au, we derive 3$\sigma$ limits of $Q$(OH) < $1.48 \times 10^{26}$ s$^{-1}$ but find no indications for [O I], $C_2$, $C_3$, or $NH_2$. From our latest X-Shooter measurements, conducted on 2025 August 21 ($r_h = 2.85$ au), we measure production rates of $\log Q(\mathrm{CN}) = 24.81 \pm 0.01$ molecules s$^{-1}$ and $\log Q(\mathrm{Ni}) = 23.30 \pm 0.07$ atoms s$^{-1}$ and characterize their evolution as the comet approaches perihelion. We observe a steep heliocentric distance scaling for the production rates $Q(\mathrm{Ni}) \propto r_h^{-7.7 \pm 1.0}$ and $Q(\mathrm{CN}) \propto r_h^{-6.7 \pm 0.2}$, and we predict an Ni–$CO_{(2)}$ correlation if the Ni I emission is driven by the carbonyl formation channel. Energetic considerations of activation barriers show that this behavior is inconsistent with the direct sublimation of canonical metal/sulfide phases and instead favors low-activation-energy release from dust—e.g., photon-stimulated desorption or mild thermolysis of metalated organics or Ni-rich nanophases, possibly including Ni–carbonyl-like complexes. These hypotheses will be testable with future coordinated ground-based and space-based monitoring, as 3I becomes more active during its continued passage through the solar system.

## 1. Introduction

### 1.1. The History of Interstellar Object Volatile Observations

Characterizing the volatile composition of interstellar objects (ISOs) passing through the solar system provides a unique window onto the chemical and physical processes operating in distant stellar systems (e.g., D. Jewitt & D. Z. Seligman 2023). ISOs preserve signatures of the chemical and physical processes operating in their natal protoplanetary disks, potentially modified by exposure to the interstellar medium. When heated by solar radiation, cometary ISOs release solids and gas due to cometary activity (e.g., 'Oumuamua ISSI Team et al. 2019).

The discovery of 1I/2017 U1 (or 'Oumuamua; hereafter, 1I) on 2017 October 19 marked the first confirmed detection of an ISO passing through our solar system. With a hyperbolic orbit (eccentricity $e \approx 1.2$) and an unusual elongated shape (aspect ratio > 5:1), 1I displayed several peculiar characteristics (K. J. Meech et al. 2017). Despite intensive observational campaigns, no coma or outgassing was directly detected around 1I, with upper limits placed on the production rates of CN, $C_2$, and $C_3$ (Q.-Z. Ye et al. 2017). However, nongravitational acceleration was observed in its trajectory, suggesting that some outgassing was occurring below the detection threshold (M. Micheli et al. 2018). The lack of observable volatiles, combined with its discovery after perihelion, when the object was faint and already receding from the Sun, limited our ability to characterize its composition (A. Fitzsimmons et al. 2018; 'Oumuamua ISSI Team et al. 2019).

---

* Based on observations collected at the European Southern Observatory under ESO programmes 115.29F3 and 115.27ZL.
[16] CATA Post-doctoral Fellow.

The second confirmed ISO, 2I/Borisov (hereafter, 2I), was discovered on 2019 August 30 (G. Borisov & Minor Planet Center 2019). Unlike 1I, this object was discovered before perihelion ($q = 2.0$ au; 2019 December 8) and exhibited a significant dust coma (J. de León et al. 2019; G. V. Borisov & B. M. Shustov 2021). Initial photometry found broadband optical colors similar to other active comets, and the discovery of this active interstellar comet allowed spectroscopic investigations of the sublimated ices. A. Fitzsimmons et al. (2019) reported the first detection of gas emitted via the near-UV emission of CN (B-X) at a heliocentric distance of $r = 2.7$ au, with a production rate of $Q(\mathrm{CN}) = (3.7 \pm 0.4) \times 10^{24}\ \mathrm{s}^{-1}$. Subsequent observations revealed additional molecular species, including $C_2$ (H. W. Lin et al. 2020), [O I] (A. J. McKay et al. 2020) and OH (Z. Xing et al. 2020), $NH_2$ (M. T. Bannister et al. 2020), and exceptionally high abundances of CO (D. Bodewits et al. 2020; M. A. Cordiner et al. 2020). Perhaps most surprisingly, P. Guzik & M. Drahus (2021) and C. Opitom et al. (2021) detected atomic nickel vapor and, in addition, C. Opitom et al. (2021) detected iron vapor in 2I, species that had just recently been discovered to be ubiquitous in the comae of solar system comets (J. Manfroid et al. 2021) and even far from the Sun. Overall, these observations demonstrated that the gas, dust, and nuclear properties of the first active ISO were broadly similar to those of solar system comets, though with some distinctive compositional characteristics.

### 1.2. Cometary Volatiles

Cometary activity is primarily driven by the sublimation of major volatiles—$H_2O$, $CO_2$, and CO—from the solar-heated nucleus. While numerous other species—such as $CH_4$, $C_3H_6$, HCN, $NH_3$, $CH_3OH$, and $CH_3CN$—are detected in cometary comae, these three dominant volatiles account for the bulk of the gas production that generates observable activity (N. Biver et al. 2024). Water ice is a major component of comets and one that dominates cometary activity and physical evolution, but CO and $CO_2$ appear to be the major "super volatiles" capable of driving activity at great solar distances, as seen in the case of Hale–Bopp (J. Crovisier et al. 1997; O. Harrington Pinto et al. 2022) and the transient brightening of 29P/Schwassmann–Wachmann 1 (M. Gunnarsson et al. 2008). The relative abundances of the different molecular species can reveal information about the formation temperature, radiation processing history, and storage conditions in the outer regions of planetary systems and interstellar space.

Spectroscopic observations of comets reveal a diverse array of volatile species with different sublimation temperatures, providing insights into their formation conditions and thermal histories (M. J. Mumma & S. B. Charnley 2011; D. Bockelée-Morvan & N. Biver 2017; K. Altwegg et al. 2019; P. Weissman et al. 2020; N. Biver et al. 2024). Highly volatile species such as CO and $CO_2$ can sublimate at temperatures as low as 25–80 K, driving activity at large heliocentric distances beyond 5 au (M. F. A'Hearn 2017). At intermediate distances (2–4 au), volatile species including HCN, $CH_3OH$, and $H_2CO$ (which sublimate at temperatures of 90–100 K) become detectable in cometary comae. Closer to the Sun, at $\lesssim 2$ au, water ice dominates the activity as surface temperatures reach 150–180 K (K. J. Meech et al. 2004).

Until recently, metallic species such as iron and nickel were thought to be released only under extreme conditions, as they are typically bound in refractory minerals with high sublimation temperatures of >700 K (G. W. Preston 1967). However, the unexpected detection of atomic nickel and iron in the cold comae of solar system comets at heliocentric distances up to 3.5 au has challenged this understanding, suggesting the existence of previously unrecognized metal-bearing molecules with low sublimation temperatures (S. J. Bromley et al. 2021; J. Manfroid et al. 2021). This diverse inventory of molecular and atomic species in cometary comae provides crucial chemical fingerprints for understanding their formation environments and subsequent evolution.

### 1.3. A New Opportunity for ISO Volatile Studies

On 2025 July 1, the Asteroid Terrestrial-impact Last Alert System (ATLAS; J. L. Tonry et al. 2018) discovered the third ISO, 3I/ATLAS,[17] at a heliocentric distance of 4.5 au, on an incoming strongly hyperbolic trajectory, having an eccentricity of $e \approx 6.14$ (L. Denneau et al. 2025; M. J. Hopkins et al. 2025; B. T. Bolin et al. 2025; D. Z. Seligman et al. 2025). The discovery occurred shortly before the beginning of the Vera C. Rubin Observatory's Legacy Survey of Space and Time (Ž. Ivezić et al. 2019), proving timely, since ISO occurrence rates remain poorly constrained given the small sample size and uncertain detection efficiency (e.g., A. Moro-Martín et al. 2009; T. Engelhardt et al. 2017; A. Do et al. 2018).

Initial observations revealed a developing coma and tail, confirming the cometary nature of 3I/ATLAS (e.g., C. O. Chandler et al. 2025; C. Opitom et al. 2025). Preliminary spectroscopic studies have shown a red reflectance spectrum (C. Opitom et al. 2025; B. T. Bolin et al. 2025; T. H. Puzia et al. 2025; D. Z. Seligman et al. 2025), possible evidence for OH emission (Z. Xing et al. 2025), and the presence of water ice (B. Yang et al. 2025). With an upper limit on its diameter of $D \lesssim 5.6$ km (D. Jewitt et al. 2025), 3I/ATLAS could potentially be larger than the previous ISOs, though its exact size remains uncertain.

What makes 3I/ATLAS particularly valuable for scientific study is the combination of its early discovery and favorable orbital parameters. Its detection several months before perihelion has allowed a comprehensive monitoring campaign to track the evolution of its activity as it approaches perihelion on 2025 October 29 at $q = 1.356$ au.[18] This extended observational timeline provides an unprecedented opportunity for detailed characterization of an extrasolar minor body throughout its journey in the inner solar system.

Here, we report the first detections of the onset of both atomic nickel (Ni I) and CN gas emission in spectroscopic observations of 3I/ATLAS obtained at the ESO Very Large Telescope (VLT) on Cerro Paranal in Chile. The observations span heliocentric distances from $r_h = 4.4$ to 2.85 au, and we examine the evolution of their production rates as the comet approaches perihelion. Detections of both molecular and atomic species in 3I/ATLAS provide important constraints

[17] https://minorplanetcenter.net/mpec/K25/K25N51.html

[18] JPL#25 data arc: 2025 May 22–2025 August 12.

**Table 1**
Journal of X-Shooter and UVES Observations

| Visit[a] | Date (UT) | Target | Exp. Time (s) | Airmass[b] | Seeing[c] (arcsec) | $r_h$ (au) | $v_h$ (km $s^{-1}$) | Δ (au) |
|---|---|---|---|---|---|---|---|---|
| X1 | 2025-07-04 06:02–07:16 | 3I/ATLAS | 3 × 900 | 1.18–1.43 | 0.69–0.81 | 4.41 | –57.63 | 3.41 |
| X2 | 2025-07-05 01:50–02:43 | 3I/ATLAS | 3 × 900 | 1.05–1.12 | 0.68–0.80 | 4.36 | –57.61 | 3.39 |
| X3 | 2025-07-17 01:44–01:44 | HD 150469 | 15 | 1.01 | 0.86 | … | … | … |
| X3 | 2025-07-17 01:55–02:46 | 3I/ATLAS | 6 × 300 | 1.00–1.02 | 0.85–0.91 | 3.98 | –57.14 | 3.08 |
| X4 | 2025-07-20 04:10–04:10 | HD 150469 | 20 | 1.25 | 0.94 | … | … | … |
| X4 | 2025-07-20 04:20–05:52 | 3I/ATLAS | 8 × 300 | 1.17–1.55 | 0.93–1.56 | 3.88 | –56.99 | 3.01 |
| X5 | 2025-07-23 02:29–03:24 | 3I/ATLAS | 4 × 600 | 1.02–1.07 | 0.84–1.06 | 3.78 | –56.83 | 2.96 |
| X5 | 2025-07-23 03:44-03:44 | HD 150469 | 20 | 1.20 | 0.74 | … | … | … |
| X6 | 2025-07-27 02:24–03:46 | 3I/ATLAS | 4 × 600 | 1.04–1.18 | 1.17–1.25 | 3.65 | –56.59 | 2.89 |
| X6 | 2025-07-27 04:11–04:11 | HD 150469 | 20 | 1.38 | 0.90 | … | … | … |
| X7 | 2025-07-30 02:24–03:46 | 3I/ATLAS | 4 × 600 | 1.07–1.24 | 0.68–0.76 | 3.55 | –56.39 | 2.84 |
| X7 | 2025-07-30 04:10–04:10 | HD 150469 | 20 | 1.44 | 0.82 | … | … | … |
| X8 | 2025-08-09 00:17–01:30 | 3I/ATLAS | 4 × 600 | 1.01–1.09 | 0.83–1.07 | 3.23 | –55.57 | 2.72 |
| X8 | 2025-08-09 02:16–02:16 | HD 150469 | 20 | 1.14 | 0.73 | … | … | … |
| U1 | 2025-08-12 23:27–00:57$^{+1d}$ | 3I/ATLAS | 3 × 1600 | 1.01–1.03 | 0.97–1.02 | 3.14 | –55.26 | 2.70 |
| X9 | 2025-08-14 01:43–03:43 | 3I/ATLAS | 4 × 600 | 1.20–1.93 | 0.92–1.12 | 3.07 | –55.03 | 2.67 |
| X9 | 2025-08-14 04:07–04:08 | HD 150469 | 20 | 1.92 | 0.67 | … | … | … |
| U2 | 2025-08-15 02:26–03:55 | 3I/ATLAS | 3 × 1600 | 1.40–1.90 | 0.83–1.23 | 3.04 | –54.91 | 2.66 |
| X10 | 2025-08-16 01:21–03:45 | 3I/ATLAS | 6 × 600 | 1.18–2.16 | 0.86–1.03 | 3.01 | –54.79 | 2.65 |
| X10 | 2025-08-16 04:09–04:09 | HD 150469 | 20 | 2.06 | 0.81 | … | … | … |
| X11 | 2025-08-21 01:17–02:49 | 3I/ATLAS | 6 × 600 | 1.26–2.07 | 0.92 | 2.85 | –54.10 | 2.62 |
| X11 | 2025-08-21 03:23–03:23 | HD 150469 | 20 | 1.73 | 0.91 | … | … | … |

**Notes.**
[a] Observing visit identifiers for X-Shooter (X1–X10) and UVES (U1–U2). Each entry corresponds to a distinct observing sequence, including both comet and solar-analog exposures.
[b] Airmass range of the target during the visit.
[c] DIMM seeing values from Paranal ambient telemetry, bracketing the visit.

on its volatile composition and enable direct comparison with both 2I and solar system comets, offering new insights into the universality of cometary material across different planetary systems.

## 2. Observations and Data Reduction

In this Letter, we make use of spectroscopic observations from X-Shooter (J. Vernet et al. 2011), on Unit Telescope 3 (Melipal), and the Ultraviolet and Visual Echelle Spectrograph (UVES; H. Dekker et al. 2000), on Unit Telescope 2 (Kueyen), of the ESO 8.2 m VLT.

To characterize the temporal evolution of 3I/ATLAS, we conduct a spectrophotometric monitoring campaign with X-Shooter under an ESO Director's Discretionary Time program (Prog. ID: 115.29F3; PI: Puzia). X-Shooter provides simultaneous wavelength coverage from the near-UV atmospheric cutoff (∼3000 Å) to the near-IR (∼24800 Å), by splitting the incoming light into three arms (UVB, VIS, and NIR), each optimized for its spectral range (J. Vernet et al. 2011). This capability delivers medium-resolution spectra across a broad wavelength range in a single visit, making it ideal for efficient ISO monitoring. In parallel, we used high-resolution spectroscopy from UVES (Prog. ID: 115.27ZL; PI: Opitom). We will present detailed analyses of each dataset in forthcoming Letters. Table 1 contains a journal of the observations. Together, these data provide measurements on narrow atomic and molecular feature fluxes and, in particular, allow us to trace the emission of Ni I and related species.

### 2.1. X-Shooter Data

In the following, we provide a description of the methods used to obtain the X-Shooter products analyzed in this work. We focus here on UVB-arm spectroscopy, while a complete reduction and analysis of all three arms will be presented in Paper I (T. H. Puzia et al. 2025, in preparation).

We designed the monitoring program for a cadence of ∼3–5 days, subject to scheduling constraints from VLTi–UT runs, technical time slots, Moon separation, weather conditions, and synchronization with other VLT instruments observing 3I/ATLAS. We observed with the 1.″6 wide slit, oriented at the parallactic angle, providing a spectral resolution $R \simeq 3200$. We did not employ nodding but used small dithers along the 11″ long slit to mitigate detector artifacts.

We reduced the raw frames with the ESO X-Shooter pipeline (v3.8.3; A. Modigliani et al. 2010) within the ESOREFLEX environment (W. Freudling et al. 2013). Standard calibration steps included bias subtraction, flat-fielding, arc lamp wavelength calibration, 2D rectification, order merging, and flux calibration. The pipeline delivered flux-calibrated, rectified, 2D slit frames.

Depending on the scientific goal, we adopted two complementary reduction and extraction approaches for 3I/ATLAS. Owing to the limited 11″slit length and the progressive increase of the comet's dust, ion, and gas coma, different strategies for sky modeling were required. Observations by W. B. Hoogendam et al. (2025) using Keck/KCWI integral field unit (IFU) spectroscopy demonstrated that the CN gas coma extends significantly beyond the Ni I emission, implying that within the 11″ X-Shooter slit, sky subtraction can lead to the self-subtraction of extended gas emission and bias line flux measurements.

For analyses focused on the dust continuum, we employed the standard ESOREFLEX sky-subtraction procedure and extracted spectra using a point-spread function (PSF)–traced aperture of 1.5 FWHM along the spatial axis. This adaptive extraction follows the centroid and FWHM of the spatial profile as a function of wavelength, thereby mitigating instrumental gradients across echelle orders and avoiding artificial color variations (F. Schönebeck et al. 2014).

In contrast, for gas emission analyses, we produced a second reduction without applying sky subtraction. Under this methodology, we then extract 1D spectra within a fixed 7″ long aperture, tracing only the comet's center. This helps in minimizing the impacts of pointing offsets and differential atmospheric refraction at high airmass, while excluding the outermost one to three spatial rows at each slit edge that are often affected by bad pixels or flat-field residuals. Furthermore, this approach simplifies the conversion of flux into the production rate. This approach preserves the total flux from the inner coma and enables consistent line measurements across visits, albeit at the cost of degradation in the continuum model, due to sky lines and continuum. For example, visit 8 (X8), taken under bright moon conditions, exhibits a strong blue continuum that hampers CN detection, whereas visit 7 shows a clear CN signal.

We measure and report dust continuum slopes using the first (sky-subtracted and PSF-traced) method, while the emission-line fluxes and derived production rates are based on the second (non-sky-subtracted and fixed-aperture-size) extraction.

From the extracted 1D spectra of each exposure, we discarded frames contaminated by field stars, identified through slit profiles and finding charts. We then averaged the remaining spectra by visit (total integration times 1800–3600 s) to produce a time series of 3I/ATLAS spectra (see Figure 1). Table 1 summarizes the observations, observing conditions, and heliocentric distances of 3I/ATLAS at the time of observation. Visit X2 was strongly affected by field star contamination in all exposures, and its continuum slope remains systematically biased.

Observations of solar analogs were obtained in short single exposures and reduced following the same procedures. The first two visits lacked dedicated solar analogs and were paired with the analog from visit X5, which was acquired under similar conditions. All subsequent visits included one or two solar-analog stars observed close in time to the target, among which HD 150469 was always one of them.

We modeled the continuum using the paired solar-analog spectrum of each visit. The model was defined as

$$F_{\text{cont}}(\lambda) = S\, R(\lambda)\, F_{\odot,\text{analog}}[\lambda(1 + v/c) + \delta\lambda], \qquad (1)$$

where $S$ is a normalization factor, $R(\lambda)$ is a polynomial reflectance function (second-order when sky subtraction is applied, seventh-order otherwise), and $F_{\odot,\text{analog}}$ is the analog spectrum shifted by the free parameters $v$ and $\delta\lambda$ to allow for residual mismatches in velocity and dispersion. We optimized the fit via rms minimization with iterative $\sigma$-clipping to exclude emission features and noise-dominated regions. While $v$ and $\delta\lambda$ act as nonphysical nuisance terms, $R(\lambda)$ describes the reflectance slope for the sky-subtracted reduction (see Section 3.5).

The high signal-to-noise ratio (S/N) of the solar-analog spectra and the low covariance among the fitted parameters render continuum model uncertainties negligible compared to the noise of the 3I/ATLAS spectra. Thus, the propagated errors in the continuum-subtracted spectra are dominated by the statistical uncertainties of the science frames. This continuum removal enables the clear identification and measurement of gas and refractory emission features in the coma, which we analyze in the following sections.

### 2.2. UVES Data

We obtained spectra of 3I/ATLAS on 2025 August 11 and 15 UT with VLT/UVES (H. Dekker et al. 2000), using two blue settings chosen to span the OH near-UV band (∼3100 Å) through the CN violet band (∼3880 Å). On the first night, we employed the 346 nm standard setting with a $1\farcs8 \times 10''$ slit, covering 3030–3880 Å and achieving a resolving power of $R \simeq 35{,}000$ while maximizing throughput. On the second night, we used the 390 nm setting with a $0\farcs6 \times 8''$ slit, covering 3260–4540 Å and reaching $R \simeq 80{,}000$ on the CN band. For each setting, we obtained three consecutive 1600 s exposures of the faint target and averaged them by night.

Table 1 lists the UVES observing dates, setups, and observational circumstances. We reduced the raw frames with the ESO UVES pipeline (v6.5.3; P. Ballester et al. 2000), including wavelength calibration, extinction correction, and flux calibration using master response curves. We then applied custom routines for spectrum extraction over the full slit length and for cosmic-ray removal. The spectra were corrected for the Doppler shift due to the comet's topocentric velocity. Finally, we removed the dust continuum using the Kurucz solar spectrum,[19] resampled to match the resolution of each setting. More details concerning the reduction methodology can be found in J. Manfroid et al. (2009).

## 3. Analysis

### 3.1. Spectral Analysis and Emission-line Identification

In both X-Shooter and UVES 3I/ATLAS spectra, we identified both atomic and molecular emission features above the underlying continuum. In X-Shooter data, to isolate these emission features, we removed the continuum by subtracting a scaled and reddened solar reference spectrum. This process revealed several emission features, most notably multiple Ni I atomic lines in the 3400–3600 Å region (beginning in visit X5 on 2025 July 23) and the CN (B-X) band around 3880 Å with detection since visit X6 (2025 July 27).

### 3.2. Production Rate Calculation Methodology

For both the molecular and atomic species detected in our spectra, we followed established methodologies to convert the measured emission-line fluxes, from the continuum-subtracted spectra, to gas production rates (J. Manfroid et al. 2021). To extract the line fluxes, for the Ni I lines we fit individual Gaussian profiles to each identified atomic line, while for CN, we approximated the asymmetric band shape with two free-fitted Gaussians in the X-Shooter data. We adopted an outflow velocity of $v = 0.85 \times r_h^{-0.5}$ km s$^{-1}$ (A. L. Cochran & D. G. Schleicher 1993) for all calculations. Due to our rectangular spectroscopic aperture ($1\farcs6 \times 7''$), we numerically integrated the spatial distribution model within the slit to derive the corresponding production rates.

[19] http://kurucz.harvard.edu/Sun.html

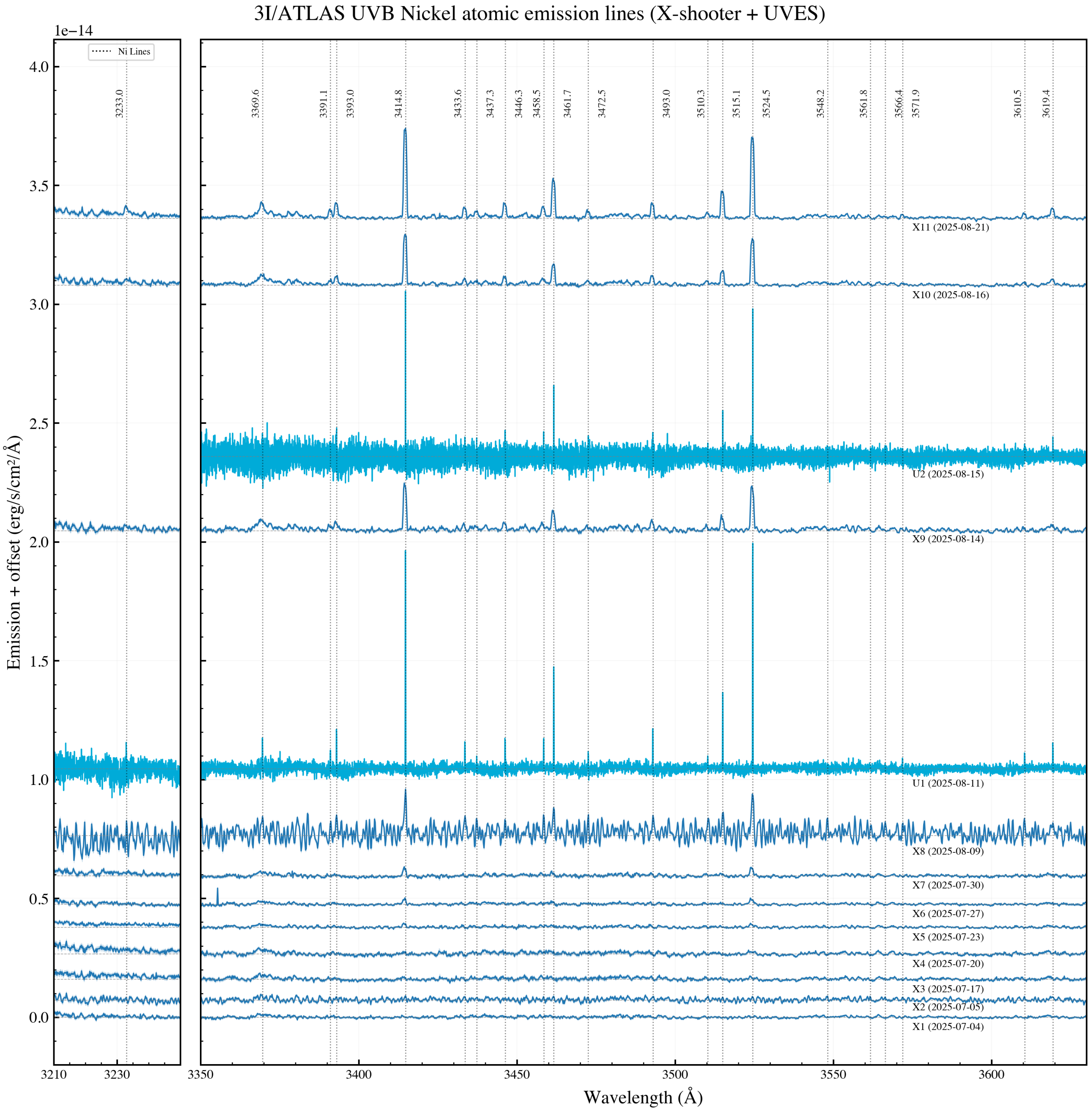

**Figure 1.** Continuum-subtracted UV/blue spectra of 3I/ATLAS showing Ni I emission over 11 VLT/X-Shooter (blue) and two VLT/UVES (cyan) visits. The spectra are vertically offset for clarity, with visit labels and dates indicated. The UVES data are scaled to match X-Shooter continuum noise levels (using visits 9–11 as references) for visualization. Left panel: 3210–3255 Å region showing selected Ni I lines. Right panel*:* 3350–3630 Å region displaying the full forest of Ni I transitions. The dotted vertical lines mark the laboratory wavelengths of the detected Ni I lines (in angstroms). The emission strengthens toward later epochs, with the first significant detection at $r_h = 3.78$ au in visit X5 (2025 July 23).

### *3.3. CN Emission*

The CN (0-0) emission band at 3880 Å was first detected when the comet was at a heliocentric distance ($r_h$) of 3.65 au in visit X6 on 2025 July 27 and also in visit U1 on 2025 August 12. Figure 2 shows the continuum-subtracted spectrum in the 3860–3930 Å region, highlighting the characteristic asymmetric profile that confirms the identification as CN. In addition, individual lines of CN are detected in higher-resolution UVES observations, adding to the confidence of the CN detection.

Following the methodology described in J. Manfroid et al. (2021), we derived CN production rates using a Haser model. We used the fluorescence efficiencies from D. G. Schleicher (2010) for the $B^2\Sigma^+$–$X^2\Sigma^+$ (0,0) band near 3880 Å, adjusted for our observing circumstances (heliocentric velocity of $-55.03$ km s$^{-1}$ at $r_h = 3.07$ au). The parent scale length of $1.3 \times 10^4$ km and daughter scale length of $2.1 \times 10^5$ km scaled $r^{-2}$ at 1 au are adopted from M. F. A'Hearn et al. (1995). The CN radical has a photodissociation lifetime that results in these characteristic scale lengths in the cometary coma. The profile distribution was integrated over the area of each of the spectrograph slits (X-Shooter: $1\farcs6 \times 7''$; UVES: 390 nm with $0\farcs6 \times 8''$ and 346 nm with $1\farcs8 \times 10''$) and compared to the observed flux to yield the production rate. For UVES visits U1 and U2, taken on 2025 August 12 and 2025 August 15, respectively, the measured CN production rates were U1: Log $Q$(CN) $= 24.23 \pm 0.30$ molecules s$^{-1}$; and U2: Log $Q$(CN) $= 24.59 \pm 0.05$ molecules s$^{-1}$. Furthermore, the coaddition of the nine brightest OH lines in the UVES 346 nm spectrum at a distance of $r_h = 3.14$ au provides us with 3$\sigma$ upper limits of $Q$(OH) $< 1.48 \times 10^{26}$ s$^{-1}$. This provides the upper limit on $Q$(CN)/$Q$(OH) $< 0.01$. For X-Shooter visit X6, taken on 2025 July 27, the measured CN-band flux of $(6.57 \pm 0.32) \times 10^{-16}$ erg s$^{-1}$ cm$^{-2}$ yielded a CN production rate of Log $Q$(CN) $= 24.11 \pm 0.02$ molecules s$^{-1}$. Table 2 lists the production rates for all visits where CN was detected and the upper limits.

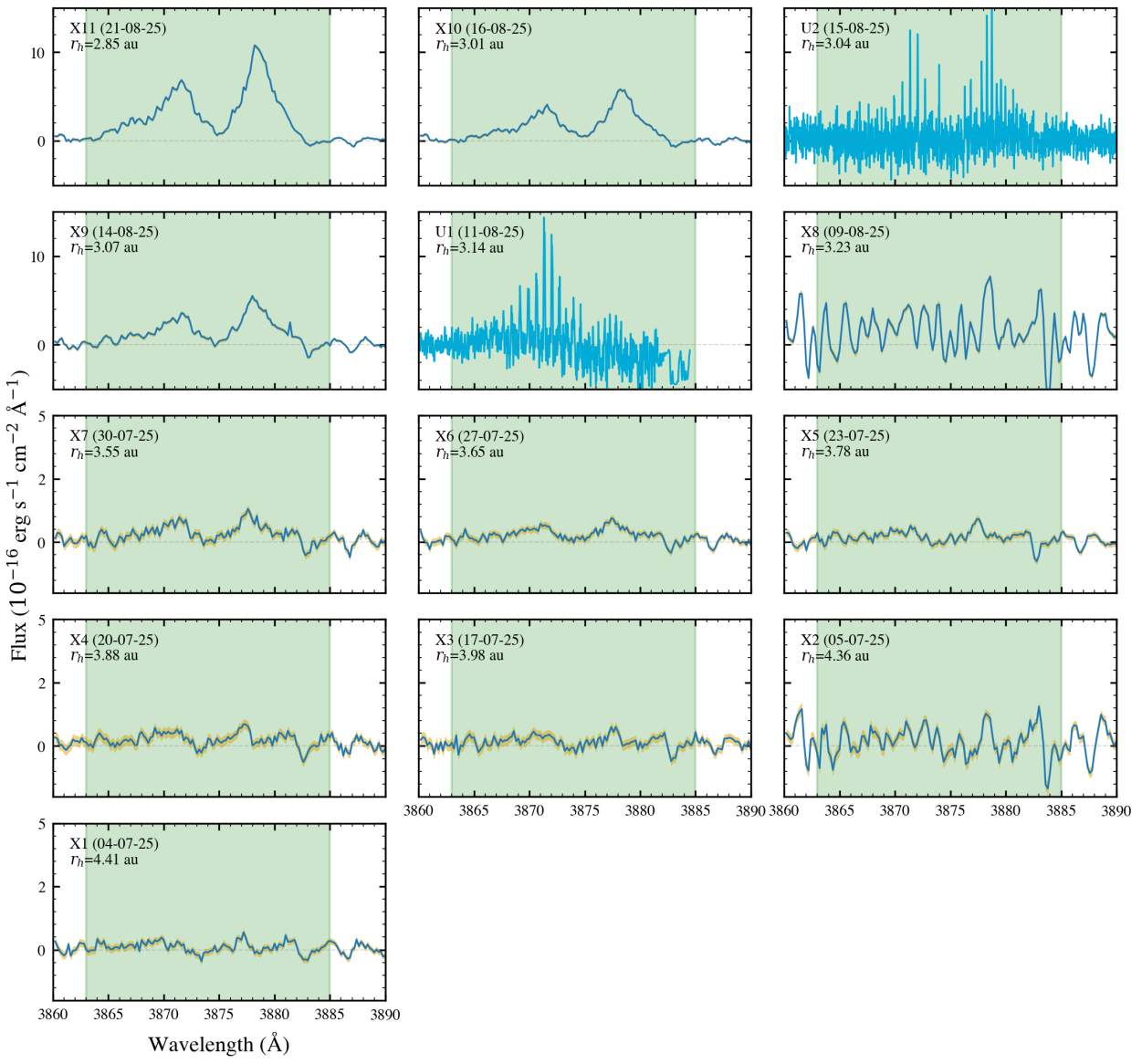

**Figure 2.** Continuum-subtracted spectra of 3I/ATLAS in the CN $B^2\Sigma^+$–$X^2\Sigma^+$(0–0) violet band region (3860–3890 Å) from a monitoring campaign spanning 2025 July 4–August 21. The grid displays 11 X-Shooter spectra (X1–X11) and two high-resolution UVES observations (U1 and U2), arranged by decreasing heliocentric distance, from top left to bottom right. Each panel indicates the observation date and heliocentric distance $r_h$ (in astronomical units). CN emission is first detected in visit X6 (2025 July 27, $r_h = 3.65$ au) and increases in strength as the comet approaches the Sun. The green shaded region marks the CN-band feature. UVES spectra show finer spectral structure compared to the lower-resolution X-Shooter observations. The shaded regions represent 1$\sigma$ uncertainties.

### 3.4. Atomic Nickel Emission

We identified 22 emission lines corresponding to Ni I transitions in our 3I/ATLAS spectra, with the strongest ones appearing starting from visit X5 on 2025 July 23. Figure 1 shows the continuum-subtracted spectrum in the 3350–3630 Å region, with the identified Ni I lines marked. The strongest lines were detected at 3414.8 and 3524.5 Å, consistent with the transitions previously observed in 2I (P. Guzik & M. Drahus 2021) and solar system comets (J. Manfroid et al. 2021).

Following the methodology of J. Manfroid et al. (2021), we computed theoretical intensities for all Ni I lines using a multilevel fluorescence model that accounts for the detailed structure of the solar spectrum. For each visit, we compared these theoretical intensities with the observed line intensities, to derive the column densities. This approach, fully detailed in J. Manfroid et al. (2021), provides a robust determination of atomic nickel abundances without relying on g-factor approximations.

For the strongest line at 3524.5 Å in visit X11, we measured a flux of $4.00 \pm 0.34 \times 10^{-15}$ erg s$^{-1}$ cm$^{-2}$, using the full fluorescence model. The spatial distribution was modeled following J. Manfroid et al. (2021), who demonstrated that metallic lines show a radiance inversely proportional to the projected distance to the nucleus, corresponding to a parent species with a short lifetime. This yielded a nickel production rate of log $Q$(Ni) $= 23.30 \pm 0.07$ atoms s$^{-1}$ for visit X11. Table 2 summarizes the derived production rates for all visits where Ni I was detected.

### 3.5. Dust Reflectance Spectrum

To characterize the dust in the coma of 3I/ATLAS in our X-Shooter observations, we refer back to the reflectance model described in Section 2.1. Although we fit a second-order polynomial for the relative reflectance, the resulting slope is nearly linear. Between 3900 and 5550 Å, the dust reflectance exhibits a reddened spectrum with a color slope of $(22 \pm 1)$%/1000 Å, which remains constant across all visits within this uncertainty. At shorter wavelengths (3200–3900 Å), the average slope is consistent with this value, but individual visits show higher scatter that we attribute primarily to the higher noise floor in this spectral region.

## 4. Discussion

The UV/blue spectral region (3200–4000 Å) shows dramatic evolution throughout our observing campaign. Figure 1 and Figure 2 present stacked spectra from all visits, clearly showing the emergence and strengthening of emission features against the dust continuum as 3I/ATLAS approaches perihelion. There is no clear trend in the evolution of the dust color across visits, suggesting that the onset of volatile activity has not produced significant changes in dust size so far.

### 4.1. Time Evolution of Ni Emission

Our spectroscopic monitoring of 3I/ATLAS provides a unique opportunity to track the evolution of gas emissions as the comet warms toward perihelion (see the thermal model in T. H. Puzia et al. 2025). Figure 3 shows the derived production rates of Ni plotted against heliocentric distance. Significant Ni I emission was first detected in visit X5 (on 2025 July 23) at a heliocentric distance of $r_h = 3.78$ au, with a production rate of log $Q$(Ni) $= 22.21 \pm 0.13$ atoms s$^{-1}$. As 3I/ATLAS approaches perihelion (2025 October 29), we observe a rapid and steady increase in nickel production, reaching log $Q$(Ni) $= 23.30 \pm 0.07$ atoms s$^{-1}$ by visit X11 at $r_h = 2.85$ au. Notably, the relative strengths of the different Ni I lines have remained consistent throughout our observing campaign so far, suggesting that the excitation conditions and

**Table 2**
Derived 3I/ATLAS Production Rates for Ni I and CN Emission

| Visit | $n_{\rm lines}$(Ni I) | $F_{\rm Ni\ I}$(3524.5Å) (erg s$^{-1}$ cm$^{-2}$) | $F_{\rm CN}$ (erg s$^{-1}$ cm$^{-2}$) | $f_{\rm Haser}$[a] | $N_{\rm total}$[a] | Production Rates log $Q$(CN) (s$^{-1}$) | Production Rates log $Q$(Ni I) (s$^{-1}$) |
|---|---|---|---|---|---|---|---|
| X5 | 4 | $2.50 \times 10^{-16}$ | $<4.50 \times 10^{-16}$ | $7.60 \times 10^{-5}$ | $6.98 \times 10^{30}$ | $< 23.99$ | $22.21 \pm 0.13$ |
| X6 | 6 | $3.17 \times 10^{-16}$ | $6.57 \times 10^{-16}$ | $7.70 \times 10^{-5}$ | $8.47 \times 10^{30}$ | $24.11 \pm 0.02$ | $22.44 \pm 0.16$ |
| X7 | 6 | $1.57 \times 10^{-16}$ | $7.68 \times 10^{-16}$ | $8.20 \times 10^{-5}$ | $8.59 \times 10^{30}$ | $24.14 \pm 0.02$ | $22.57 \pm 0.12$ |
| X8 | 10 | $1.56 \times 10^{-15}$ | $3.72 \times 10^{-15}$ | $1.10 \times 10^{-4}$ | $2.54 \times 10^{31}$ | $24.72 \pm 0.04$ | $23.12 \pm 0.05$ |
| U1 | 20 | … | $1.95 \times 10^{-15}$ | $1.76 \times 10^{-4}$ | $7.47 \times 10^{30}$ | $24.23 \pm 0.30$ | $22.82 \pm 0.06$ |
| X9 | 14 | $2.20 \times 10^{-15}$ | $3.14 \times 10^{-15}$ | $1.20 \times 10^{-4}$ | $1.63 \times 10^{31}$ | $24.58 \pm 0.01$ | $23.09 \pm 0.07$ |
| U2 | 14 | … | $1.46 \times 10^{-15}$ | $5.71 \times 10^{-5}$ | $1.56 \times 10^{31}$ | $24.59 \pm 0.05$ | $22.74 \pm 0.08$ |
| X10 | 16 | $2.34 \times 10^{-15}$ | $3.62 \times 10^{-15}$ | $1.30 \times 10^{-4}$ | $1.67 \times 10^{31}$ | $24.62 \pm 0.01$ | $23.15 \pm 0.08$ |
| X11 | 18 | $4.00 \times 10^{-15}$ | $6.75 \times 10^{-15}$ | $1.60 \times 10^{-4}$ | $2.32 \times 10^{31}$ | $24.81 \pm 0.01$ | $23.30 \pm 0.07$ |

**Notes.**
[a] Haser model fraction in slit—see M. F. A'Hearn et al. (1995) for further details.
[b] Total CN molecules in coma.

fluorescence mechanism did not change significantly over this period.

The Ni production rate follows a power-law dependence on heliocentric distance of the form Q(Ni) $\propto r_h^{-7.7\pm1.0}$, which is steeper than any purely radiative scaling. The equilibrium surface temperature of an airless body scales as $T \propto r_h^{-0.5}$, and between $r_h \simeq 3.9$ and 2.85 au this raises $T$ by ∼15%–20%. Photon-driven rates—e.g., via UV desorption or resonance fluorescence per parent—scale at most with the solar flux, i.e., $\Phi_\odot \propto r_h^{-2}$. Our 3I/ATLAS measurements yield a scaling far steeper than both $r_h^{-0.5}$ (temperature) and $r_h^{-2}$ (insolation), implying that the supply of Ni atoms to the coma is growing superlinearly as the object approaches the Sun. This, in turn, indicates that the rate-limiting step is not the availability of photons, nor a linear response to temperature increase, but may instead likely be a temperature-activated and/or threshold process that is being switched on.

If the parent of Ni is released by a process with the Arrhenius form $Q \propto \exp(-E_a/kT)$ and $T \propto r_h^{-0.5}$, then

$$\frac{d\ln Q}{d\ln r_h} = \frac{E_a}{kT^2}\left(-\frac{T}{2}\right) = -\frac{E_a}{2kT}, \qquad (2)$$

where $E_a$ is the activation energy and $k$ is the Boltzmann constant. If, over a limited range, $Q(r_h)$ behaves like a power law, then with the previous result we obtain

$$Q \propto r_h^{-n} \;\Rightarrow\; \frac{d\ln Q}{d\ln r_h} = -n \;\Rightarrow\; n = \frac{E_a}{2kT}. \qquad (3)$$

At moderate temperatures $T \simeq 150$–200 K and a measured slope $n \simeq 7.7$, we estimate an activation energy of $E_a = 2kTn \approx 0.20$–0.27 eV or 19–27 kJ mol$^{-1}$, which is typical for desorption or diffusion barriers for weakly bound species on ices or organics (e.g., K. Furuya et al. 2022) and crucially does not imply metal sublimation (e.g., P. D. Feldman et al. 2004). This points to Ni-bearing organometallics (metal carbonyls; see, e.g., P. Guzik & M. Drahus 2021; D. Hutsemékers et al. 2021; J. Manfroid et al. 2021) or grain-surface complexes (metal polycyclic aromatic hydrocarbons or PAHs) as candidates (A. Klotz et al. 1995; I. Ristorcelli & A. Klotz 1997; S. J. Bromley et al. 2021), rather than direct sputtering or simple photon-stimulated desorption (e.g., T. Hama & N. Watanabe 2013; H. M. Cuppen et al. 2017).

If Ni I is predominantly released from refractory grains via thermal/UV processes in the coma, then once volatile outgassing ($CO_2$/$H_2O$) exceeds the cohesion threshold, the lifted dust mass can rise very steeply with decreasing $r_h$ (e.g., B. Gundlach & J. Blum 2015; B. Gundlach et al. 2015). Other interesting potential mechanisms are the temperature crossing of the amorphous-to-crystalline $H_2O$ transition at $T \simeq 120$–150 K, the polymerization of ices, and/or the devolatilization of surface organics, which can trigger both gas releases and fragmentation, effectively increasing the active area and the stock of Ni-bearing carriers (e.g., A. Bar-Nun & D. Laufer 2003; R. Miles 2016; M. Womack et al. 2017; C. A. Schambeau et al. 2019; S. Faggi et al. 2024). Furthermore, changing the subsolar latitude or erosion can expose fresh terrains enriched in the Ni carrier, compounding any thermal activation.

The power-law index $n \simeq 7.7$ is steeper than typically observed for volatile species in solar system comets (M. F. A'Hearn et al. 1995; M. M. Knight & D. G. Schleicher 2013; C. Opitom et al. 2015a, 2015b; M. R. Combi et al. 2019), which mostly show $r_h^{-1}$ to $r_h^{-6}$ rate dependencies, albeit at shorter heliocentric distances. We note that inbound slopes are typically steeper than post-perihelion slopes. The steeper dependence observed for nickel may indicate a distinct production onset mechanism, possibly related to its hypothesized origin from a short-lived parent molecule, such as nickel carbonyl (J. Manfroid et al. 2021). Future monitoring observations are advised to test for correlations with dust tracers (continuum, Na, etc.) or with Q($H_2O$), Q($CO_2$), and Q(CO), where a strong correlation—particularly one steeper than linear proportionality—would support dust-entrainment and/or phase-transition drivers.

### *4.2. Time Evolution of CN Emission*

3I/ATLAS's CN emission was first measured in our X-Shooter spectra on 2025 July 27 (visit X6) at a heliocentric distance of 3.65 au, with a production rate of $\log Q(\mathrm{CN}) = 24.11 \pm 0.02$ molecules s$^{-1}$. Figure 4 shows the derived production rates of CN plotted against heliocentric distance. The later appearance of CN as compared to Ni I suggests different thermal thresholds for the release of their respective parent molecules, with the putative nickel parent $Ni(CO)_4$ sublimating at lower temperatures than HCN ($\lesssim$76–79 K; see, e.g., N. Fray & B. Schmitt 2009), the likely parent of CN (N. Fray et al. 2005). We recognize that other effects might also be at play, such as differences in detection sensitivity, relative abundances, and the stratification of parent molecules at different depths within the nucleus (see the thermal evolution model in T. H. Puzia et al. 2025).

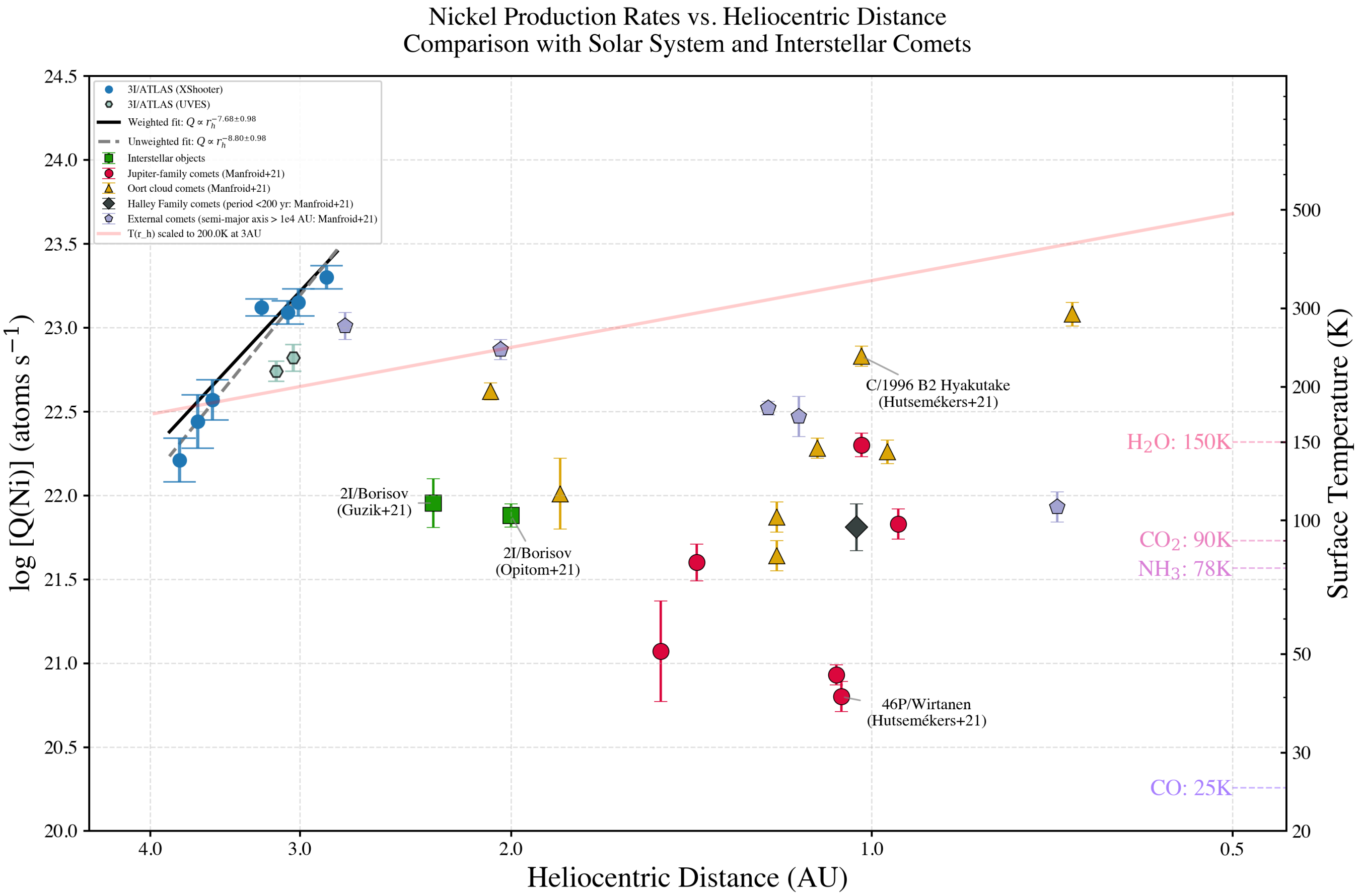

**Figure 3.** Nickel production versus heliocentric distance: measurements from VLT/X-Shooter (blue circles) and VLT/UVES (teal hexagons) obtained in this work are shown for comet 3I/ATLAS—the black solid and gray dashed curves are power-law fits to the 3I data only, giving $Q(\mathrm{Ni}) \propto r_h^{-7.7\pm1.0}$ (weighted) and $Q(\mathrm{Ni}) \propto r_h^{-8.8\pm1.0}$ (unweighted), respectively. The colored symbols compare with solar system comets compiled from the literature, grouped by dynamical class as in the legend; the green squares mark measurements for 2I. The left axis gives $\log_{10}[Q(\mathrm{Ni})/\mathrm{atoms\ s^{-1}}]$ and the bottom axis gives in log scale the heliocentric distance $r_h$ (au). The right axis shows an approximate equilibrium surface temperature $T(r_h)$—orange line, scaled to 200 K at $r_h$ = 3 au—to guide the eye (see the text for details). The horizontal dashed lines indicate sublimation thresholds for CO, $NH_3$, $CO_2$, and $H_2O$. Other comet references: P. Guzik & M. Drahus (2021), D. Hutsemékers et al. (2021), J. Manfroid et al. (2021), and C. Opitom et al. (2021).

Given the lower-resolution spectrum of X-Shooter compared to UVES when resolving the CN band, we cannot robustly constrain the heliocentric distance dependence of the Q(CN) production rate. However, the available data suggest a power-law index of approximately $r_h^{-6.7\,\pm\,0.2}$, which is steep but within the range observed for solar system comets (−10 to −1.5; see M. F. A'Hearn et al. 1995). In the latter fit, we excluded visit X8, because its S/N was too low (sky-limited).

We note that small discrepancies between the X-Shooter and UVES measurements are expected, given the use of solar-analog calibration in the former (E. Tatsumi et al. 2022) and the limited spectral resolution of X-Shooter in resolving the CN band. Nevertheless, the X-Shooter results remain self-consistent across visits, as all reductions relied on the same solar-analog reference star observed under similar conditions to 3I/ATLAS (see Table 1).

L. E. Salazar Manzano et al. (2025) reported their first CN detection on 2025 August 10 and production rates $Q(\mathrm{CN}) < (5.8 \pm 1.9) \times 10^{24}\ \mathrm{s}^{-1}$, increasing to $Q(\mathrm{CN}) \approx (4.82 \pm 0.34) \times 10^{24}\ \mathrm{s}^{-1}$ to $(7.17 \pm 0.38) \times 10^{24}\ \mathrm{s}^{-1}$ thereafter, during the 2025 August 12–17 period. Narrowband photoelectric photometry obtained by D. Schleicher (2025) on 2025 August 19 using a 97″ aperture derived $Q(\mathrm{CN}) \approx (7.23^{+1.28}_{-1.09}) \times 10^{24}\ \mathrm{s}^{-1}$. Production rates from these studies are also plotted in Figure 4. They are in good agreement with the X-Shooter results for the given heliocentric distance.

### 4.3. Comparison with Solar System Comets and Other ISOs

The detection of atomic nickel and CN in 3I/ATLAS allows us to place this ISO (A. G. Taylor & D. Z. Seligman 2025) in context with both 2I and the diverse population of solar system comets. Figure 3 presents our nickel production rate measurements alongside those from other comets, revealing distinct patterns across different comet populations. The Ni production rates we measure for 3I/ATLAS, ranging from $\log Q(\mathrm{Ni}) = 22.21 \pm 0.13$ atoms $\mathrm{s}^{-1}$ at 3.78 au to $\log Q(\mathrm{Ni}) = 23.30 \pm 0.07$ atoms $\mathrm{s}^{-1}$ at 2.85 au, extend the heliocentric distance range for nickel detections in comets, providing new constraints on production mechanisms. For context, scaling our measurement at 2.85 au using a canonical $r_h^{-2}$ relationship would yield $Q(\mathrm{Ni})|_{1\,\mathrm{au}} \approx 1.62 \times 10^{24}\ \mathrm{s}^{-1}$, already surpassing the values reported for solar system benchmarks such as the notably bright comet C/1996 B2 (Hyakutake) with $Q(\mathrm{Ni}) \simeq 10^{22}$–$10^{23}\ \mathrm{s^{-1}}$ (S. J. Bromley et al. 2021; D. Hutsemékers et al. 2021).

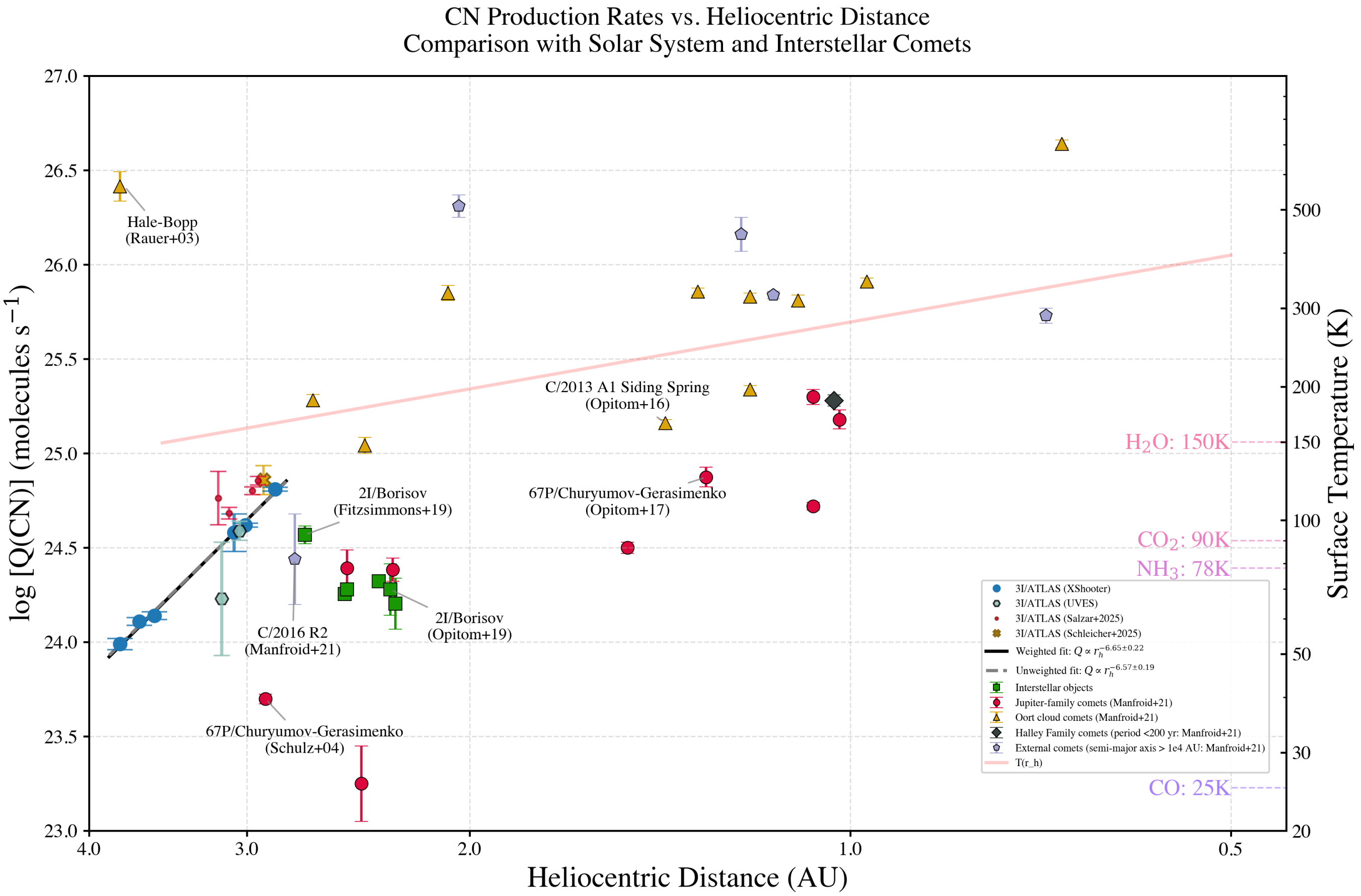

**Figure 4.** CN production versus heliocentric distance: measurements from VLT/X-Shooter (blue circles) and VLT/UVES (teal hexagons) obtained in this work are shown for 3I/ATLAS—the black solid and gray dashed curves are power-law fits to the 3I data only, giving $Q(\mathrm{CN}) \propto r_h^{-6.7\pm0.2}$ (weighted) and $Q(\mathrm{CN}) \propto r_h^{-6.6\pm0.2}$ (unweighted), respectively. The colored symbols compare with solar system comets compiled from the literature, grouped by dynamical class as in the legend; the green squares mark measurements for 2I. The left axis gives $\log_{10}[Q(\mathrm{CN})/\mathrm{molecules\ s^{-1}}]$ and the bottom axis gives in log scale the heliocentric distance $r_h$ (au). The right axis shows an approximate equilibrium surface temperature $T(r_h)$. The horizontal dashed lines indicate sublimation thresholds for CO, $NH_3$, $CO_2$, and $H_2O$. Other comet references: H. Rauer et al. (2003), R. Schulz et al. (2004), C. Opitom et al. (2016), C. Opitom et al. (2017), A. Fitzsimmons et al. (2019), C. Opitom et al. (2019), and J. Manfroid et al. (2021).

In Figure 3, the external and Oort cloud comets (OCCs) show higher Ni production rates ($10^{21.5}$–$10^{23.1}$ atoms $\mathrm{s^{-1}}$) compared to Jupiter-family comets (JFCs; $10^{20.8}$–$10^{22.3}$ atoms $\mathrm{s^{-1}}$) at similar heliocentric distances. These are based on data from J. Manfroid et al. (2021), who demonstrated that JFCs display larger variance in Ni I/Fe I ratios compared to OCCs. Our analysis of their data yields an F-statistic of 4.3 with a $p$-value of approximately 7%, indicating that JFCs show significantly more heterogeneity in their Ni I/Fe I ratios compared to OCCs. This may suggest that JFCs formed under more diverse conditions in the outer solar nebula. However, it should be noted that OCCs also tend to show higher activity levels in general, not restricted to Ni, potentially due to a reduction of the active surface area as comets become more processed. In addition, observational biases (e.g., the smaller number of JFCs observed at larger heliocentric distances) can complicate direct comparisons between the two families. With these caveats in mind, 3I/ATLAS's relatively high nickel production rates at our closest approach (2.85 au) are more consistent with external and OCC levels, providing potential clues to their formation environment and/or evolutionary history.

Compared to 2I, which showed $Q(\mathrm{Ni}) = (0.9 \pm 0.3) \times 10^{22}$ atoms $\mathrm{s^{-1}}$ at 2.3 au (P. Guzik & M. Drahus 2021), 3I/ATLAS exhibits higher nickel production at comparable distances. The average Ni I/Fe I ratio in solar system comets is approximately 15.5 times higher than solar values (J. Manfroid et al. 2021). This preferential nickel enhancement in cometary volatile reservoirs may extend to ISOs as well. However, since no Fe I lines were detected in our X-Shooter and UVES data, additional observations of 3I/ATLAS are needed to determine its Ni I/Fe I ratio and make a direct comparison.

The sequential detection of Ni I (3.78 au) and CN (3.65 au) in 3I/ATLAS differs from the reported observations of 2I, where CN was detected at 2.7 au (A. Fitzsimmons et al. 2019) and metals at 2.3 au (P. Guzik & M. Drahus 2021). While these detection sequences might suggest compositional differences, they could also reflect observational biases, such as sensitivity or different observing circumstances (e.g, pre-/post-perihelion positioning). Nevertheless, the detection of both species in ISOs strengthens the broader hypothesis that metal atoms in cometary comae may originate from molecules with low sublimation temperatures, possibly metal carbonyls (S. J. Bromley et al. 2021; J. Manfroid et al. 2021).

Our CN detection in visit X11 on 2025 August 21 with $Q(\mathrm{CN}) = 6.5 \times 10^{24}$ molecules $\mathrm{s^{-1}}$ at 2.85 au can be compared with 2I's $Q(\mathrm{CN}) = 3.7 \pm 0.4 \times 10^{24}$ molecules $s^{-1}$ at 2.7 au (A. Fitzsimmons et al. 2019). Among solar system comets, CN production spans a wide range, from $10^{23}$ to $10^{27}$ molecules $s^{-1}$, depending on comet size and heliocentric distance (M. F. A'Hearn et al. 1995; U. Fink 2009; A. L. Cochran et al. 2012). Representative examples include C/2013 A1 (Siding Spring), with $Q(\mathrm{CN}) = 1.4 \times 10^{25}$ molecules $s^{-1}$ at $r_h = 1.40$ au (C. Opitom et al. 2016), and 67P/Churyumov–Gerasimenko, with peak production of $Q(\mathrm{CN}) = 7.5 \pm 0.9 \times 10^{24}$ molecules $s^{-1}$ approximately two weeks after perihelion (C. Opitom et al. 2017).

The detection of similar metal species and comparable production rates in both ISOs observed to date suggests that there may be commonalities in the processes governing protoplanetary disk chemistry across different stellar systems. With an estimated age older than the solar system (M. J. Hopkins et al. 2025; A. G. Taylor & D. Z. Seligman 2025), 3I/ATLAS provides a window onto organometallic chemistry from a significantly different and likely lower-metallicity epoch of galactic evolution compared to our 4.6 billion yr old environment, potentially revealing how these processes operate in planetary systems formed much earlier than our own.

### *4.4. Implications for Origin and Formation Environment*

The detection of Ni I emission in a cold coma with absent Fe I implies short-lived, low-temperature, Ni I-bearing parent molecules that photodissociate close to the nucleus. Such detections were previously made in 2I near $r_h \approx 2.3$ au (P. Guzik & M. Drahus 2021). Converting our measured Ni I line fluxes to production rates requires modeling the spatial distribution in the coma. Following J. Manfroid et al. (2021), we assume a $1/r$ radial profile, characteristic of a short-lived parent, implying a parent lifetime of a few times $10^2$–$10^3$ s at 1 au and an Ni I production rate $\sim 10^{22}$ $s^{-1}$ ($\sim 2 \times 10^{-3}$ of OH)—i.e., a species volatile enough to be released at $\lesssim$200 K and photolabile enough to yield atoms near their source (e.g., D. Hutsemékers et al. 2021; J. Manfroid et al. 2021). The absence of Fe I under similar conditions demands either: (a) an Ni-specific parent; and/or (b) a release/processing pathway that favors Ni over Fe at low $T$, modulo differences in fluorescence efficiency.

#### *4.4.1. Metal Carbonyls*

Nickel tetracarbonyl, $Ni(CO)_4$, is exceptionally volatile[20] and undergoes ultrafast photodissociation under UV irradiation,[21] producing Ni I atoms and CO on subpicosecond-to-picosecond timescales, precisely the behavior needed for a short-lived Ni I parent in the inner coma (R. R. Ford 1970).[22]

Carbonyls form by the direct reaction of $Ni^0$ with CO (the Mond process; L. Mond et al. 1890), efficient at modest temperatures ($\sim$320–350 K). In astrophysical settings, trace $Ni^0$ or freshly reduced Ni on mineral surfaces in CO-rich pores/ices can carbonylate during transient warm-up in disks or parent bodies, then become trapped and later released by sublimation or photodesorption at $\lesssim$200 K. The sequential coordination of CO to adsorbed Ni I atoms on cold grain/ice surfaces can be stabilized by the ISO surface (third-body effect), forming $Ni(CO)_4$, which subsequently desorbs and photolyzes in the coma. Such cold and UV-bright conditions provide a natural Ni I/Fe I asymmetry. For example, $Fe(CO)_5$ is considerably less stable than $Ni(CO)_4$ in the presence of water and oxidants, and hydrolysis/solvolysis selectively suppresses Fe carbonyls where trace $Ni(CO)_4$ can persist. In water-rich cometary comae/ices, this can yield Ni I without Fe I (R. R. Ford 1970), which we currently observe in 3I/ATLAS at $r_h \simeq 2.85$ au.

Beyond hydrolysis, the redox partitioning of refractory phases in primitive solids favors Fe I in oxidized/sulfidized minerals, whereas minor $Ni^0$ metal grains (and Ni I atoms produced by irradiation/sputtering) can remain locally available for carbonylation. Even trace $Ni(CO)_4$ inventories suffice to explain rates of $\sim 10^{22}$ Ni $s^{-1}$ (microgram $s^{-1}$ levels). If this scenario generally applies to 3I/ATLAS's surface conditions, Ni I production should correlate more strongly with $Q(\mathrm{CO})$ than with $Q(H_2O)$. The fundamental CO bands of $Ni(CO)_4$ fall near 4.7–5.1 $\mu$m, and deep mid-IR spectroscopy with JWST could place direct limits on $Ni(CO)_4$ abundances.

On 2025 August 6, JWST/NIRSpec IFU observations (0.6–5.3 $\mu$m) revealed a $CO_2/H_2O$ mixing ratio of $7.6 \pm 0.3$ in 3I's coma at $r_h = 3.32$ au (M. A. Cordiner et al. 2025). In 2025 mid-August, SPHEREx found a bright resolved $CO_2$ gas coma with $Q(CO_2) \approx (9.4 \pm 2.8) \times 10^{26}$ $s^{-1}$, while placing $3\sigma$ limits of $Q(H_2O) \lesssim 1.5 \times 10^{26}$ and $Q(\mathrm{CO}) \lesssim 2.8 \times 10^{26}$ $s^{-1}$ (C. M. Lisse et al. 2025). These measurements imply $CO_2$-dominated activity at $r_h \gtrsim 3.2$ au and evaporative cooling that can keep grain/nucleus temperatures near $\sim$120 K, suppressing $H_2O$ vapor while still efficiently lofting icy/dusty grains. These conditions naturally favor rapid dust entrainment with decreasing $r_h$ and the release of low-activation-energy Ni carriers. In combination, these measurements favor the metal carbonyl formation channel.

#### *4.4.2. Metal PAH or Fullerene Complexes*

An alternative formation channel is transition metals that attach to PAHs (A. G. G. M. Tielens 2008) nearly barrierlessly, via radiative association on grains or in cold gas. Such complexes photofragment efficiently, potentially releasing neutral Ni I atoms plus PAH radicals, yielding a short-lived parent with nucleus-centered Ni I. PAHs and related carbonaceous macromolecules are widespread in the interstellar medium (P. Ehrenfreund & S. B. Charnley 2000; B. T. Draine 2003; E. Herbst & E. F. van Dishoeck 2009; J. P. Williams & L. A. Cieza 2011). Metal PAH (and metal fullerene) adducts have been discussed in interstellar chemistry and have been explicitly proposed as parents for cometary Ni and Fe (S. J. Bromley et al. 2021).

Differences in binding motifs and photostability between Ni–PAH and Fe–PAH complexes can bias survivability. If Fe analogs oxidize or hydrate more readily in $H_2O$ ice, Ni complexes could dominate the volatile organometallic inventory, yielding Ni without Fe in the inner coma, as observed here in 3I/ATLAS. If Ni parents are PAH-related, trends of Ni/Fe with classical carbon chain diagnostics (e.g., $C_2$/CN classes) observed across comets should extend to 3I/ATLAS. In fact, we observe the onset of CN but no Swan bands at 3 au. Any association with the 3.3, 6.2, 7.7, 8.6, or 11.2 $\mu$m aromatic

[20] $Ni(CO)_4$ has a sublimation temperature $\sim$74–82 K compared to $Fe(CO)_5$, which sublimates at $\sim$97–108 K.

[21] Note that we are near solar maximum in 2025 August when our observations were taken.

[22] See the standard data compilations and photokinetic studies in the NIST Chemistry WebBook thermochemical data: https://webbook.nist.gov/chemistry/.

infrared bands (A. G. G. M. Tielens 2008) would be circumstantial support for metal PAHs.

#### 4.4.3. Hybrid Scenario: Direct Release from Ni-bearing Minerals Plus In-coma Carbonylation

UV sputtering or thermochemical erosion can liberate Ni atoms from Ni sulfides (e.g., pentlandite) or rare metal grains. In a CO-rich microenvironment, they rapidly carbonylate on/near grain surfaces to $Ni(CO)_4$, which then photodissociates to Ni I with a short parent scale length. This preserves the carbonyl-like photophysics without requiring prestored $Ni(CO)_4$ in the nucleus (R. R. Ford 1970; D. Brownlee et al. 2006; A. Bardyn et al. 2017).

For comparison, solar wind sputtering has been observed to release the refractory species Na, K, Si, and Ca in the coma of comet 67P/Churyumov–Gerasimenko at $r_h \approx 3$ au (P. Wurz et al. 2015). They found that sputtering signals diminish as comets approach the Sun, because the solar wind becomes absorbed by the thickening gas layer above the surface and can no longer reach the nucleus. In contrast, we observe Ni I emission strengthening (not weakening) as 3I/ATLAS approaches perihelion from 3.9 to 2.85 au. Furthermore, our derived activation energy of $E_a \simeq 0.20$–0.27 eV (Section 4.1) suggests a thermally activated release mechanism rather than the momentum transfer process of sputtering, which is independent of surface temperature. These considerations strongly disfavor sputtering as the source of Ni I in 3I/ATLAS.

It will be interesting to see as future data are collected whether Ni I tracks both the dust release (the source of Ni) and CO gas. Spatially, Ni should be strongly nucleus-centered (parent lifetime $\ll 10^3$ km at 1 au), with possible enhancements where dust jets intersect CO-rich outflows (P. Guzik & M. Drahus 2021), yielding anisotropies.

#### 4.4.4. The Need for Future Diagnostics

A short-lived parent producing Ni at $\sim 10^{-3}$% of OH is naturally delivered by Ni–carbonyl photolysis and, secondarily, by Ni–PAH photofragmentation. The nondetection of Fe I in 3I/ATLAS to $r_h \simeq 3$ au is chemically reasonable if $Fe(CO)_5$ is destroyed or never forms in water-rich, mildly oxidizing conditions while $Ni(CO)_4$ survives (R. R. Ford 1970; P. Guzik & M. Drahus 2021; J. Manfroid et al. 2021). Continued spectroscopic monitoring of 3I/ATLAS is crucial for tracking Ni I flux versus $Q$(CO) and versus dust continuum, as carbonyl-mediated mechanisms predict stronger linkage to CO than to $H_2O$ (P. Guzik & M. Drahus 2021). Furthermore, if Ni I is present with Fe I below limits consistent with $Fe(CO)_5$ quenching, the carbonyl asymmetry is strongly favored (D. Hutsemékers et al. 2021; J. Manfroid et al. 2021). However, remote-sensing caveats on inferring bulk refractory abundances from coma measurements have to be kept in mind (M. F. A'Hearn 2017), especially in the currently active solar cycle environment.[23]

## 5. Conclusions and Outlook

We have presented UV/blue VLT/X-Shooter and UVES spectroscopy of the interstellar comet 3I/ATLAS. At $r_h \simeq 4.4$–2.85 au, the coma is dust-dominated and red, while classical daughter species are undetected: [O I], $C_2$, $C_3$, or $NH_2$. We detect the onset of CN emission in both X-Shooter visit X6 (2025 July 27) and UVES visit U1 (2025 August 12), with production rates of $\log Q(\mathrm{CN}) = 24.11 \pm 0.02$ molecules s$^{-1}$ and $\log Q(\mathrm{CN}) = 24.23 \pm 0.30$ molecules s$^{-1}$, respectively. Against this volatile–quiet backdrop, we also report the detection of Ni I emission lines but no Fe lines with both instruments. Taken together, these results imply that nickel is being released efficiently at low equilibrium temperatures.

This is difficult to reconcile with the direct sublimation of canonical meteoritic metal or sulfide phases and instead points to low-activation-energy carriers and/or photochemical release from organics on (or within) grains. Candidate pathways include: (a) the formation and rapid dissociation of volatile Ni–carbonyl-like complexes in CO-rich microenvironments; (b) the photo-/thermodesorption of Ni from metalated organics (PAHs/fullerenes); or (c) the selective liberation from fine Ni-rich nanophases produced by aqueous or space-weathering processes in the natal system. Each route predicts measurably different spatial and temporal behaviors.

If 3I/ATLAS continues to exhibit Ni without Fe through perihelion, it will constitute the first clear case where interstellar cometary metal emission is decoupled from classical refractory release. That outcome would argue for a distinct low-temperature organometallic (or nanophase) pathway for Ni in extrasolar comets and could open a new window onto how disk chemistry, metallicity, and irradiation history imprint on planetesimal microphysics. While the parent star of 3I/ATLAS is likely to be a metal-poor relative to other ISO progenitor stars, it is unlikely to be even a factor of 2 less metal-rich than the Sun (M. J. Hopkins et al. 2025; A. G. Taylor & D. Z. Seligman 2025), meaning that there is no tension between the inferred age of 3I/ATLAS and the presence of an iron peak element like Ni.

Regardless of which scenario prevails, 3I/ATLAS offers a decisive time-critical experiment linking metal emission to volatile activation and grain physics in an ISO. The measurements outlined above will turn nickel from a curiosity into a calibrated tracer of both parent chemistry and Galactic provenance, setting the standard for rapid-response ISO spectroscopy in the Rubin Observatory and ESO Extremely Large Telescope era.

## Acknowledgments

We are grateful to the ESO Garching and Paranal staff for their extraordinary support during the implementation and execution of the observations. In particular, we thank Mario van den Ancker, Cédric Ledoux, Diego Parraguez, Luca Sbordone, Felipe Gaete, Francesca Lucertini, Cecilia Bustos, Rob van Holstein, Rodrigo Palominos, Celia Desgrange, Rodrigo Romero, and Lorena Faundez. We acknowledge support from the National Agency for Research and Development (ANID) grants CATA-Basal FB210003 and Beca de Doctorado Nacional (R.R. and J.P.C.). The authors would like to thank the anonymous referee for constructive comments that improved our manuscript. This research has made use of data and/or services provided by the International Astronomical Union's Minor Planet Center, the NASA Jet Propulsion Laboratory's Horizons System, available at https://ssd.jpl.nasa.gov/horizons/, and the NASA/IPAC Extragalactic Database, which is funded by the National Aeronautics and Space Administration and operated by the California Institute of Technology.

[23] https://www.swpc.noaa.gov/products/wsa-enlil-solar-wind-prediction

The views expressed in this Letter are those of the authors and do not reflect the official policy or position of the U.S. Naval Academy, the Department of the Navy, the Department of Defense, or the U.S. Government.

*Facilities:* VLT:Kueyen.

*Software:* astropy (Astropy Collaboration et al. 2013, 2018, 2022), matplotlib (J. D. Hunter 2007).

## ORCID iDs

Rohan Rahatgaonkar https://orcid.org/0000-0002-5350-0282
Juan Pablo Carvajal https://orcid.org/0000-0001-6584-7104
Thomas H. Puzia https://orcid.org/0000-0003-0350-7061
Baltasar Luco https://orcid.org/0009-0000-5806-5550
Emmanuel Jehin https://orcid.org/0000-0001-8923-488X
Cyrielle Opitom https://orcid.org/0000-0002-9298-7484
Jean Manfroid https://orcid.org/0000-0002-6930-2205
K. Aravind https://orcid.org/0000-0002-8328-5667
Michaël Marsset https://orcid.org/0000-0001-8617-2425
Bin Yang https://orcid.org/0000-0002-5033-9593
Laura Buchanan https://orcid.org/0000-0002-8032-4528
Wesley C. Fraser https://orcid.org/0000-0001-6680-6558
John Forbes https://orcid.org/0000-0002-1975-4449
Michele Bannister https://orcid.org/0000-0003-3257-4490
Dennis Bodewits https://orcid.org/0000-0002-2668-7248
Bryce T. Bolin https://orcid.org/0000-0002-4950-6323
Matthew Belyakov https://orcid.org/0000-0003-4778-6170
Matthew M. Knight https://orcid.org/0000-0003-2781-6897
Colin Snodgrass https://orcid.org/0000-0001-9328-2905
Rosemary Dorsey https://orcid.org/0000-0002-8910-1021
Léa Ferellec https://orcid.org/0000-0002-2088-3889
Fiorangela La Forgia https://orcid.org/0000-0003-3924-1867
Manuela Lippi https://orcid.org/0000-0001-9185-878X
Brian Murphy https://orcid.org/0000-0002-8137-5132
Prasanta K. Nayak https://orcid.org/0000-0002-4638-1035
Mathieu Vander Donckt https://orcid.org/0000-0002-2380-6889